\documentclass[conference, a4paper]{IEEEtran}
\IEEEoverridecommandlockouts
\usepackage{amsmath,amssymb,amsfonts}
\usepackage{algorithmic}
\usepackage{graphicx}
\usepackage{textcomp}
\usepackage{xcolor}
\usepackage{multirow}
\usepackage{balance}
\usepackage{epstopdf}

\usepackage{subcaption}
\usepackage{siunitx}
\DeclareSIUnit{\dbm}{dBm}
\DeclareSIUnit{\second}{s}

\usepackage{comment}
\usepackage{algorithm}
\usepackage{graphicx}

\usepackage{bm}
\usepackage{enumitem}
\usepackage{siunitx}
\DeclareSIUnit{\belmilliwatt}{Bm}
\DeclareSIUnit{\dBm}{\deci\belmilliwatt}
\usepackage{soul}
\soulregister\cite7
\soulregister\citeauthor7
\soulregister\textcite7
\soulregister\ref7
\soulregister\SI7
\soulregister\si7
\soulregister\ac7
\soulregister\acp7
\soulregister\acf7
\soulregister\o7
\soulregister\decibelm7
\soulregister\meter7
\usepackage[nolist]{acronym}

\usepackage[backend=biber,style=ieee,  sortcites=true,isbn=false, doi=false, url=false,mincitenames=1, maxcitenames=1,maxbibnames=3,hyperref=false]{biblatex}
\usepackage{booktabs}

\def\BibTeX{{\rm B\kern-.05em{\sc i\kern-.025em b}\kern-.08em
    T\kern-.1667em\lower.7ex\hbox{E}\kern-.125emX}}

\begin{document}
\title{
A Collaborative Approach Using Neural Networks for BLE-RSS Lateration-Based Indoor Positioning
}


\author{\IEEEauthorblockN{%
Pavel Pascacio%
\IEEEauthorrefmark{1}\textsuperscript{,}\IEEEauthorrefmark{2}, %
Joaquín Torres-Sospedra%
\IEEEauthorrefmark{3}, %
Sven Casteleyn%
\IEEEauthorrefmark{1}
and Elena Simona Lohan%
\IEEEauthorrefmark{2}
}

\thanks{Corresponding Author: J. Torres-Sospedra (\texttt{info@jtorr.es})}
\thanks{This work was supported by the European Union’s Horizon 2020 Research and Innovation programme under the Marie Sklodowska Curie grant agreements No. $813278$ (A-WEAR: A network for dynamic wearable applications with privacy constraints, {http://www.a-wear.eu/})  and No.~$101023072$ (ORIENTATE: Low-cost Reliable Indoor Positioning in Smart Factories, {http://orientate.dsi.uminho.pt}).}

\IEEEauthorblockA{\IEEEauthorrefmark{1}\textit{Institute of New Imaging Technologies}, \textit{Universitat Jaume I}, Castellón, Spain}
\IEEEauthorblockA{\IEEEauthorrefmark{2}\textit{Electrical Engineering Unit}, \textit{Tampere University}, Tampere, Finland}
\IEEEauthorblockA{\IEEEauthorrefmark{3}\textit{Algoritmi Research Centre}, \textit{Universidade do Minho}, Guimarães, Portugal}
}

\maketitle
\begin{abstract}
In daily life, mobile and wearable devices with high computing power, together with anchors deployed in indoor environments, form a common solution for the increasing demands for indoor location-based services. Within the technologies and methods currently in use for indoor localization, the approaches that rely on Bluetooth Low Energy (BLE) anchors, Received Signal Strength (RSS), and lateration are among the most popular, mainly because of their cheap and easy deployment and accessible infrastructure by a variety of devices. Nevertheless, such BLE- and RSS-based indoor positioning systems are prone to inaccuracies, mostly due to signal fluctuations, poor quantity of anchors deployed in the environment, and/or inappropriate anchor distributions, as well as mobile device hardware variability. In this paper, we address these issues by using a collaborative indoor positioning approach, which exploits neighboring devices as additional anchors in an extended positioning network. The collaborating devices' information (i.e., estimated positions and BLE-RSS) is processed using a multilayer perceptron (MLP) neural network by taking into account the device specificity in order to estimate the relative distances. After this, the lateration is applied to collaboratively estimate the device position. Finally, the stand-alone and collaborative position estimates are combined, providing the final position estimate for each device. The experimental results demonstrate that the proposed collaborative approach outperforms the stand-alone lateration method in terms of positioning accuracy.
\end{abstract}

\begin{IEEEkeywords}
Collaborative Indoor Positioning, Multilayer Perceptron, Received Signal Strength, Bluetooth Low Energy 
\end{IEEEkeywords}

\begin{acronym}[XXX] 
\acro{ann}[ANN]{Artificial Neural Network}
\acro{bs}[BS]{base Station}
\acro{ble}[BLE] {Bluetooth Low Energy}
\acro{cdf}[CDF]{Cumulative distribution Function}
\acro{cips}[CIPS]{Collaborative Indoor Positioning System}
\acro{crlb}[CRLB]{Cram\'er Rao Lower Bound}
\acro{ecdf}[ECDF] {Empirical Cumulative Distribution Function}
\acro{gdop}[GDOP]{Geometric Dilution of Precision}
\acro{gnss}[GNSS]{Global Navigation Satellite System}
\acro{gt}[GT]{Ground-Truth} 
 \acro{ips}[IPS]{Indoor Positioning System}
 \acro{lbs}[LBS]{location-based service}
 \acro{ldpl}[LDPL]{Logarithm Distance Path Loss}
 \acrodef{lstm}[LSTM]{Long Short-Term Memory}
 \acro{los}[LOS]{Line-of-sight}
\acro{mad}[MAD]{Median Absolute Deviation}

\acro{mlp}[MLP]{Multilayer Perceptron}

\acro{mse}[MSE]{Mean Square Error}

 \acro{nlos}[NLOS]{Non-line-of-sight}
\acro{pca}[PCA]{Principal Component Analysis}
\acro{rf}[RF]{Radio Frequency}
\acro{rmse}[RMSE]{Root Mean Square Error}
 \acro{rss}[RSS]{Received Signal Strength}
 \acro{rssi}[RSSI]{Received Signal Strength Indicator}
\acro{siso}[SISO]{Single Input Single Output}
\acro{tof}[ToF]{Time of Flight}
\acro{twr}[TWR]{Two-way Ranging}
\acro{uwb}[UWB]{Ultra-wide band}
\end{acronym}

\section{Introduction}
 Nowadays, there is a growing demand for indoor \acp{lbs} which are able to ubiquitously and accurately provide a user's position relative to the surrounding environment in an inexpensive manner \cite{basiri2017indoor}. Examples of \ac{lbs} include navigation and location-aware and tracking services \cite{s21061950,he2021user, cheema2018indoor}. To this end, the pervasiveness of mobile and wearable devices with high computing power, together with growing deployment of anchors in indoor environments, play an important role. Among the large variety of technologies and methods used, solutions relying on \ac{ble} and \ac{rss}  are very popular, as they provide a cheap and easy deployment \cite{jeon2018ble,li2018indoor,yang2009indoor,ARANDA2022117095}.

Although the feasibility of using \ac{ble}--\ac{rss} anchors/devices for proximity detection applications has been demonstrated \cite{montanari2017study, ng2019compressive}, their implementation for accurate positioning applications is still an open issue. Due to their relatively straightforward implementation, lateration algorithms are one of the most explored solutions to enhance positioning accuracy. Unfortunately, they suffer from accuracy degradation mainly due to four factors:
\begin{enumerate}
    \item  Anchor deployment: a low quantity of anchors deployed in the environment and/or  the inappropriate anchor distribution increase not only the chance of \ac{nlos} conditions but also the locations inside the operational area with insufficient anchor coverage for reliable lateration, reducing the overall positioning accuracy dramatically.
    \item Unstable \ac{ble} signal strength: \ac{rss} of \ac{rf} signals are prone to fluctuate due to environmental conditions (geometries, temperature,  crowded environments), building materials and presence of obstacles or people between transmitter and receptor devices (\ac{nlos} conditions) \cite{7346778,huh2017indoor,Flueratoru2021}. Such dynamic fluctuations are often difficult to predict and compensate for, and cause position estimation errors when used for positioning. In fact, the \ac{ble} signal propagation in indoor environments has not been properly modeled yet in order to estimate the relative distance between transmitter and receiver using smartphones.
    \item  Hardware  heterogeneity: the \ac{rss} signature depends on the transmitters and receivers involved \cite{10.1145/1023720.1023728}. The diversity of devices involved in localization -- both anchors and smartphones -- comes with variability in the hardware-specific parameters, such as antenna gains, and thus in the \ac{ble} received signal strength~\cite{jin2013ssd,Flueratoru2021}. Furthermore, the transmission power and transmission period  preset in each device -- either by the device owner or manufacturer -- presents variations due to tolerance in the design of antennas or in the \ac{rf} circuit components, even in similar devices. 
\item Software limitations:
the energy-saving modes and execution task priorities prescribed by the mobile devices' operating systems may alter the transmission/reception period of \ac{ble} signals, rendering these devices non-deterministic for the \ac{ble} broadcasting.
\end{enumerate}

In this article, we aim to enhance the positioning accuracy of the stand-alone lateration method applied to an adverse indoor scenario that copes with the first, second, and third issues mentioned above,  namely the lack of sufficient beacons or problematic beacon distribution with the associated presence of large \ac{nlos} areas, the lack of appropriate signal propagation models to deal with fluctuating signal strength in unpredictable indoor conditions, and the heterogeneity among mobile devices typically present in any scenario. We do so by proposing a collaborative indoor position system which exploits the neighboring mobile devices as additional anchors in an extended, ad-hoc positioning network. The collaborating devices' information (i.e., their stand-alone estimated position and \ac{ble}--\ac{rss}) is exchanged and processed using a neural network model (rather than the traditional path-loss models) to estimate relative distances, after which lateration is applied to estimate the device's collaborative position. Unlike the path-loss models, which estimate the distance primarily based on signal attenuation formulas, the neural network model aims to relate the distance to \ac{rss} patterns, pairwise device position, identification of devices (transmitter and receiver), and  variation of \ac{rss} measurements due to the devices' hardware heterogeneity. The collaborative algorithm takes mobile device variations into account by considering, for each device, a \ac{ble} reference measurement at one meter. Finally, a midpoint line algorithm is used to combine the collaborative and the non-collaborative position estimates to provide the final position estimates of each collaborative device. The key contributions of this paper are summarized as follows:

\begin{itemize}
	\item We present a collaborative indoor position approach using \ac{mlp} \ac{ann}  to enhance \ac{ble}--\ac{rss} lateration-based indoor positioning.
	 \item We present a  \ac{mlp} \ac{ann}   model  to estimate  relative distances between collaborative devices in order to provide a most reliable distance estimation. The \ac{mlp} model provides a better modeling of the \ac{nlos} and \ac{los} conditions of the environment with respect to path-loss-based models, as well as, the characteristic inherent of each receiving device. Additionally, the proposed model avoids setting the parameters needed in path-loss-based models.
	\item We experimentally  demonstrate, using a scenario with a poor quantity and inappropriate distribution of anchors, and rich \ac{nlos} conditions, that our collaborative approach outperforms the stand-alone lateration approach with respect to positioning accuracy in such conditions. 
\end{itemize}

\section{Related Work}
\label{sec:related_work}
The design of reliable \acf{ips}, which provide a balance between position accuracy and low-cost and easy deployment, is one of the main concerns in the design of \acfp{lbs} \cite{basiri2017indoor}. To meet this concern,  diverse solutions have been proposed. In this section, we focus on those fields relevant for this work, namely lateration, artificial neural network, and collaborative \ac{ips}.

\subsection{Lateration}
Lateration approaches estimated the target position by considering the distance between the target and multiple reference anchors (minimum three)\cite{dag2018received}. The distance is the result of the conversion of \ac{rss} measured between reference anchors and target~\cite{dag2018received,cengiz2020comprehensive}. Free space path, two-ray ground, and logarithmic  path-loss are some of the most used models to convert \ac{rss} into distance~\cite{miao2018accurate,katirciouglu2011comparing,qiu2016ble}. 

In the context of indoor positioning, it has been shown in literature that having more (even redundant) anchors yield better positioning accuracy --independent of the lateration approach used-- and least-square lateration performs better than traditional lateration. For example, \textcite{dag2018received} proposed an \ac{rss}-based lateration algorithm, which was implemented using a least-squares method and required redundant anchors deployment. The distances between anchors and target were estimated with the \ac{ldpl} model, whose parameters were computed statistically. The experiments were conducted in a \SI[output-product = \times]{6 x 6}{\meter} area, considering 196 measurement points collected with a single device in complete \ac{los}. The authors conclude that adding least-squares to lateration reduces the error by $67.6\%$ to \SI{2.8}{\meter}, and including redundant anchors reduces the error to \SI{1.35}{\meter}. \textcite{cengiz2020comprehensive} also proposed a least-square lateration approach and performed an accuracy error analysis using simulated environments. The analysis included the level of Gaussian noise, number of anchors, and size of the area.
The results were compared with a traditional lateration, concluding that increasing the density of anchors deployed improves positioning accuracy. 
The proposed method significantly improved lateration in scenarios with a strong noise component (high Gaussian noise, $\sigma = 5$), reducing the error from \SIrange{4.5}{1.9}{\meter}, \SIrange{5.9}{3.9}{\meter} and  \SIrange{18}{7.5}{\meter} for \SI[product-units=power]{6 x 6}{\meter}, \SI[product-units=power]{12 x 12}{\meter} and \SI[product-units=power]{24 x 24}{\meter} test areas respectively.

In line with literature, our method uses least-square-based method for lateration. Moreover, our approach aims to increase the amount of anchors, yet in contrast to previous works that deploy additional fixed anchors in the environment, we dynamically extend the positioning network using the devices of the users in a collaborative setting. 

\subsection{Artificial neural network}

\acfp{ann} are commonly used to self-learn the relationship between data provided to resolve complex problems \cite{wu2020artificial}. Specifically, in indoor positioning they are used to provide robustness against noise and interference that affect the positioning accuracy \cite{nessa2020survey}. For example, \textcite{wu2020artificial} presented a path-loss model based on \acf{mlp} neural networks for wireless communication networks. For that purpose they considered several models using input vectors with fifteen features of the environment. The authors conclude that the introduction of environment features in the \ac{mlp} model input and adding two hidden layers (with 5 and 10 neurons) provide an accurate and efficient path-loss prediction. On the other hand, \textcite{elbes2019indoor} proposed an indoor localization approach based on Wi-Fi fingerprinting and \ac{lstm} neural networks for position estimation. The experiments were carried out in \SI[product-units=power]{54 x 32}{\meter} test L-shape area with ad-hoc Wi-Fi anchors deployed and a single smartphone to scan the signals. The \ac{lstm} model consist of $42$ input features, $2$ outputs, $300$ neurons, learning rate $\lambda = 0.2$, drop error, and $250$ epochs. The results reported an average error of around 1 \si{\meter}.

In contrast to related work, our \ac{mlp} architecture is intended for use with heterogeneous mobile devices to estimate the relative distance between neighboring devices in a collaborative approach. To this end, our \ac{mlp} architecture exploits the usual short distance between neighboring devices to reduce the dimensionality of features and, consequently, obtain a lightweight neural network architecture that is feasible to implement on mobile devices. Specifically, the \ac{mlp} architecture relates the distance to four main elements of BLE-RSS collaborative approaches. The elements used for the features of the architecture are the \ac{rss} patterns, pairwise device position, identification of devices (transmitter and receiver), and variation of \ac{rss} measurements due to devices' hardware heterogeneity.

\subsection{Collaborative indoor positioning systems}

Collaborative \acp{ips} aim to enhance positioning of devices/users by using wireless communication technologies, both for relative distance estimation and information exchange between them~\cite{pascacio2021collaborative}. 
 Within the collaborative \acp{ips}, those relying on \ac{ble}-\ac{rss} for the non-collaborative and collaborative phase, as our approach, are scarce (for a comprehensive overview of collaborative \acp{ips}, we refer to our recent systematic literature review~\cite{pascacio2021collaborative}). One of the most representative ones is proposed by~\textcite{qiu2016ble}, which presented a collaborative \ac{ips} based on inertial sensing, adaptive multi-lateration, and a mobile encountering approach. All these sources of data were combined using a particle-filter model to locate a target.  
 
The author conclude that the accuracy improvements are approximately $28.19\%$ in \ac{los} and $19.99\%$ in \ac{nlos} when using adaptive ranging device-specific parameter with respect to the none adaptive approach.

In contrast to what has been done in the literature, we do not rely on the \ac{ldpl} model to estimate the relative distance between a receiver (RX) and a transmitter (TX). Instead, we use a neural network to process the RSS value depending on the transmitter and receiver location and receiver calibration. 

 \subsection{Overall summary}
 From the above-mentioned systems, we can summarize that the lateration methods considerably decrease the positioning accuracy as the density of anchors in the area decreases and the noise signal increases. \acp{ann} can mitigate the effect of the environment on the position and distance estimation, as well as reduce the accuracy error. Additionally, the implementation of collaborative approaches can help to reduce the positioning error of individual estimations. To the best of our knowledge,  this is the first \ac{ble}--\ac{rss} indoor positioning system that implements a collaborative approach based on \acp{ann}.

\section{System Overview}
\label{sec:model}

Under simulations or in ideal testbed conditions, i.e. scenarios featuring a rich amount and well distributed anchors in the environment and \ac{los} conditions, RSS-based lateration approaches have provided high accuracy with positioning errors within the range of \SIrange{0.8}{1.3}{\meter}~\cite{dag2018received,cengiz2020comprehensive}. Nevertheless, in real environments (e.g., offices, stores, libraries) the positioning accuracy decreases dramatically. Real environments are characterized by large \ac{nlos} areas and anchor deployments that may not be exclusively designed for positioning tasks. In addition, each environment requires a dedicated modeling of the \ac{rf} signal propagation to compensate for environment-specific conditions and accurately convert \ac{rss} into distance.

In this work, we propose a collaborative \ac{ips} using \acp{ann} to enhance \ac{ble}--\ac{rss} lateration-based methods. Our approach tackles three main sources of positioning inaccuracy: mitigating the inaccuracy due to \ac{nlos} areas and inadequate deployment of anchors; reducing the impact of a inappropriate signal propagation modeling; and minimizing the impact of device heterogeneity. Potential application scenarios of our approach is indoor positioning in buildings with a poor infrastructure deployment, but with a moderate amount of mobile devices (people) inside and where the latency for users/devices positioning is moderate (e.g., tracking library assistants in libraries, patients in hospitals, personnel in government offices, among others).

Our collaborative approach is divided in three phases which are described in this section. The first phase is devoted to registering devices used in the collaborative approach, the second phase consists of the stand-alone (non-collaborative) lateration algorithm for position estimation of each device/user, and the last phase is devoted to collaboratively enhance the position of the target device/user. 

\subsection{Registering a device for collaborative positioning}
\label{sec:registering}
The registration procedure aims to identify a mobile device and record an \ac{rss} reference value, to be used in the collaborative algorithm. To this end, a fixed reference anchor $(BX)$ emitting \ac{ble} advertisements is placed at an entry point to the environment. Each new user is then asked to position himself at 1 meter from the anchor (marked on the floor) in \ac{los}, and measure and record the \ac{rss} value (i.e., $RSS^{1m}_{RX}(BX)$). This lightweight calibration procedure is performed only once for each (new) device. To ease this procedure, the calibration could be integrated with the electronic locking system of the main doors, calibrating the device every time the user's smartphone is used to unlock a door. 

\subsection{Stand-alone phase}
\label{stand-alone}
The stand-alone phase is devoted to estimating each initial individual device/user's position considering the \ac{rss} and the \ac{gt} coordinates of \ac{ble} anchors deployed in the environment. For reference, in this article this position estimation is performed using a lateration method based on the \ac{ldpl} model, a minimization problem, and the Levenberg-Marquardt algorithm for weighted nonlinear least-squares to solve the minimization problem.

The \ac{ldpl} model expresses the relation between \ac{rss} and distance, as described by Eq.~\eqref{eqn:pathloss}  \cite{qureshi2018analysis}: \begin{equation}
\label{eqn:pathloss}
RSS^{d}_{RX}(TX) = RSS^{1m}_{RX}(BX) - 10\:\eta \:log \left(\frac{d}{d_0}\right)\textbf{}
\end{equation}
Where $RSS^{d}_{RX}(TX)$ is the \ac{rss} measured at a distance $d$ between transmitter ($TX$), which can be a \ac{ble} beacon or a smartphone, and receiver ($RX$); $RSS^{1m}_{RX}(BX)$ is the \ac{rss} measured at \SI{1}{\meter}; and $\eta$ is the path-loss attenuation factor.

Mathematically, the lateration approach is expressed by Eq.~\eqref{eqn:minproblem} as a minimization of the sum of squared errors between the measured distances ($d_m$) and hypothetical ones ($g_m(\b{x})$), based on the unknown target position, $g_m(\b{x})$, which is denoted by Eq.~\eqref{eqn:gm} 
\cite{zhou2012multilateration}. \begin{equation}
\label{eqn:minproblem}
\min_{\b{x}} \displaystyle\sum_{m=1}^M \left( g_m(\b{x})-d_m \right)^2
\end{equation}
\begin{equation}
\label{eqn:gm}
g_m(\b{x})\triangleq \sqrt{\left(x-bx_m \right)^2 - \left(y-by_m \right)^2}
\end{equation}
Where, $m=\{1,2...M\}$ are the number of \ac{ble} anchors deployed; $\b{x}=\{x,y\}$ are the device/user's unknown coordinates; and $\{bx_m,by_m\}$ the \ac{gt} coordinates of \ac{ble} anchors. 

Algorithm \ref{alg:one} shows the pseudo-code for the stand-alone lateration approach, whose workflow is summarized as follows:

\begin{itemize}
    \item \textbf{1\textsuperscript{st} step:} Collect  the \ac{rss} from \ac{ble} anchors ($RSS^{d}_{RX}(TX)$) within a time window ($tw$) of  \SI{60}{\second}, by discarding those not belonging to the deployed  scenario (input data for Algorithm~\ref{alg:one});
    
    \item \textbf{2\textsuperscript{nd} step:} Group the \ac{rss} readings by anchor removing the outlier values. We consider the values falling out of 25 and 75 percentiles (lines 1--2 in Algorithm~\ref{alg:one}) as outliers;
    
     \item \textbf{3\textsuperscript{rd} step:} Apply the average operator to the \ac{rss} values of each anchor, in order to get one averaged \ac{rss} value per anchor (line 3 in Algorithm~\ref{alg:one});
     
     \item \textbf{4\textsuperscript{rd} step:} Select "strong reference \ac{ble} anchors", defined as the anchors with averaged \ac{rss} equal or greater than a predefined threshold (lines 4--8 in Algorithm~\ref{alg:one});

     \item \textbf{5\textsuperscript{th} step:} Estimate the relative distances of selected reference \ac{ble} anchors to the device/user's target position, by using the LDPL model, which is expressed by eq.\eqref{eqn:pathloss}, and their \ac{rss} values (line 9 in Algorithm~\ref{alg:one});
    
    \item \textbf{6\textsuperscript{th} step:} Estimate the device/user's position, $\hat{\textbf{P}}^{Lat1}$, using the Levenberg-Marquardt Weighted Least Squares (L-MWLS) lateration method to fit the Euclidean Distance model. Henceforth, $Lat1$ will refer to the L-MWLS lateration method. The input data to fit the model are the distances estimated in the 5\textsuperscript{th} step, the weights and the \ac{gt} of the \ac{ble} anchors. The weight value for every \ac{ble} anchor is computed as the inverse of its distance square with respect to the device/user;
    
    \item \textbf{7\textsuperscript{th} step:} Share the estimated position, $\hat{\textbf{P}}^{Lat1}(RX) = [\hat{P}^{Lat1}_x(RX),\hat{P}^{Lat1}_y(RX)]$.
\end{itemize}

We set the $threshold$ value to \SI{-83}{\dbm} and the path-loss attenuation factor to $\eta = 2.1$, as we previously used in~\cite{pascacio2021lateration} 
 and which are in phase to values in literature~\cite{dag2018received}. Each device involved in the collaborative positioning model will have its own value for $RSS^{1m}_{RX}(BX)$ calculated from the data collected in the device registration phase (see Table~\ref{rssat1mtable}).%

\begin{algorithm}[!htb]
 \caption{Stand-alone Lateration}
 \label{alg:one}
 \begin{algorithmic}[1]
 \small
 \renewcommand{\algorithmicrequire}{\textbf{Input:}}
 \renewcommand{\algorithmicensure}{\textbf{Output:}}
  \REQUIRE Deployed anchors information collected within a time window $tw$: $RSS^{d}_{RX}(TX)$ values and GT
  \REQUIRE LDPL: $\eta = 2.1$ and $RSS^{1m}_{RX}(BX)$ 
  \REQUIRE $threshold$ 
  
  \ENSURE Estimated device/user position $\hat{\textbf{P}}^{Lat1}(RX)$

\STATE Group the $RSS^{d}_{RX}(TX)$ values by beacon
\STATE Remove $RSS^{d}_{RX}(TX)$ outliers values of each group
\STATE Average $RSS^{d}_{RX}(TX)$ values of each group : $\overline{RSS^{d}_{RX}(TX)}$
\FOR {$i\leftarrow$ 1 \textbf{to} number of $\overline{SS^{d}_{RX}(TX)}(i)$}
 \IF {($\overline{SS^{d}_{RX}(TX)}(i) \ge treshold $)}
  \STATE Include $i$-th anchors to reference anchors set ($ref_{anchorset}$)
  \ENDIF
  \ENDFOR
  \STATE Estimate the distances between anchors of $ref_{anchorset}$ and the device/user position using Eq.\ref{eqn:pathloss},  $\eta$ and $RSS^{1m}_{RX}(BX)$
 \STATE Estimate the device/user position ($\hat{\textbf{P}}^{Lat1}(RX)$) using the Levenberg-Marquardt Weighted Least Squares  method 
 \STATE \textbf{ Share the estimated device/user position ($\hat{\textbf{P}}^{Lat1}(RX) = [\hat{P}^{Lat1}_x(RX),\hat{P}^{Lat1}_y(RX)]$)}
 \end{algorithmic}
 \end{algorithm}

\subsection{Collaborative phase}

The collaborative phase uses the neighboring mobile devices' information, which include  the measured \ac{ble}-\acp{rss}, the estimated positions for the receiver  
and transmitter devices, and the RSS at \SI{1}{\meter} of the receiver to collaboratively estimate the device/user's position, and ultimately provide the final position estimate based on the combined collaborative and stand-alone estimates, aiming to improve the latter. Each device/user, which exchanges information and collaborates with others, acts as an additional anchor, creating an extended ad-hoc positioning network, hereby improving the coverage of the original anchor network.

The collaborative phase relies on four main elements: 1) information exchanged between devices; 2) estimation of relative distances using a \acf{mlp} neural network; 3) a lateration algorithm to collaboratively estimate the target device/user's position; and  4) a method to combine the stand-alone and collaborative position estimations. 

\subsubsection{Information exchanged between devices/users}
Currently, the operating systems of mobile and wearable devices permit broadcasting \ac{ble} advertisements. This feature allows them to act as \ac{ble} anchors and share information, enabling the development of collaborative positioning systems~\cite{huh2016bluetooth}. The information used in our collaborative positioning system includes: the RSS received at receiver from transmitter ($RSS_{RX}(TX)$), the \ac{rss} at \SI{1}{\meter} ($RSS^{1m}_{RX}(BX)$) of the receiver obtained after registering the device and the position of each collaborative device/user estimated by the stand-alone lateration ($\hat{P}_{x}^{Lat_1}(TX) $,$\hat{P}_{y}^{Lat_1}(TX) $,$\hat{P}_{x}^{Lat_1}(RX) $,$\hat{P}_{y}^{Lat_1}(RX)$). 

All these collected data are used as inputs for the neural network presented in next subsection and the collaborative algorithm presented in Algorithm \ref{alg:two}. 

\subsubsection{\acf{mlp} neural network and relative distances}

\begin{figure}[!h]
    \centering
  \includegraphics[width=1\columnwidth,trim={2.5cm 2.5cm 2.1cm 2cm},clip]{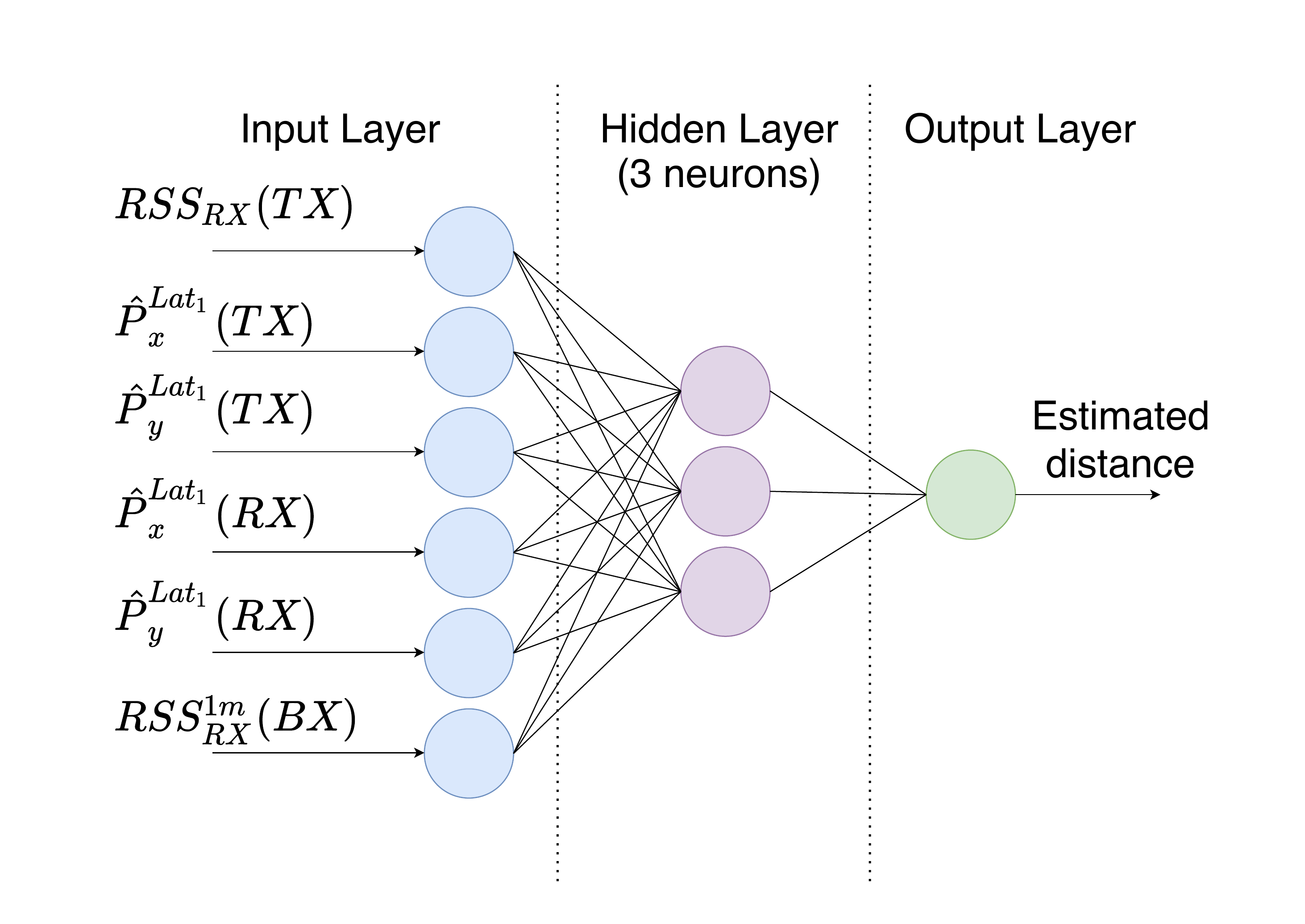}
  \caption{\ac{mlp} neural network model architecture. Used to estimate relative pairwise distance between target and neighboring devices/users.}
  \label{fig:Ann_model}
\end{figure}

The \ac{mlp} neural network model's architecture used to estimated the relative distances is presented in Figure~\ref{fig:Ann_model}. Its architecture consists of 6 input layers, 1 hidden layer with 3 neurons, and one output layer. The activation function used is the hyperbolic tangent. In the training, the scaled conjugate gradient back propagation was used as training function and $50$ epochs were considered. Further information about the selection of hyperparameters  of the \ac{mlp} neural network is provided in Section~\ref{sec:experiments}.
The 6 inputs that feed the \ac{mlp} neural network model correspond to the exchanged information between each pair (target device--neighbor device) and the output is the estimated distance of that specific pair. The input are: the $RSS_{RX}(TX)$, which corresponds to the \ac{rss} of the \ac{ble} advertisement sent by the transmitter $TX$ and measured at the receiver $RX$;  the $x$ and $y$ coordinates of the transmitter (neighbor device/user) position estimated by the stand-alone lateration algorithm ($\hat{P}_{x}^{Lat_1}(TX) $,$\hat{P}_{y}^{Lat_1}(TX) $); and the $x$ and $y$ coordinates of the receiver (target device/user) position estimated by the stand-alone lateration algorithm ($\hat{P}_{x}^{Lat_1}(RX) $,$\hat{P}_{y}^{Lat_1}(RX)$).

Finally, the $RSS^{1m}_{RX}(BX)$ is the \ac{rss} value measured at \SI{1}{\meter} of distance between transmitter and receiver assigned at each mobile device by phase one (see Table~\ref{rssat1mtable}). 

\subsubsection{Collaborative lateration algorithm}

The lateration algorithm used in the collaborative phase is the same algorithm used in the stand-alone phase (see Section \ref{stand-alone}). The main two differences are within the data used to perform the lateration.

\begin{itemize}
    \item First, rather than fixed \ac{ble} beacons in the environment with well-known position, the anchors considered in the collaborative lateration algorithm are the neighboring devices/users, of which only the position estimated by the stand-alone lateration is known. 

\item Second, the relative distances between the target device/user and the neighboring devices/users is computed using a \ac{mlp} neural network model instead of the \ac{ldpl} model.
\end{itemize}
\subsubsection{Combining stand-alone and collaborative estimated positions}
Once the position of the target device is collaboratively estimated using neighbor devices/users as anchors, it is combined with the stand-alone estimation to come to a final estimated position. In this article, we used a midpoint line algorithm, as described in Eq.~\eqref{eqn:joinequation}.%
\begin{align}
\label{eqn:joinequation}
\hat{\hat{P}}_{x}(RX) &=
\frac{\hat{P}^{lat_1}_{x}(RX)
+
\hat{P}^{lat_2}_{x}(RX)}{2}\\
\hat{\hat{P}}_{y}(RX) &=
\frac{\hat{P}^{lat_1}_{y}(RX)
+
\hat{P}^{lat_2}_{y}(RX)}{2}
\end{align}
Where $\hat{\hat{\textbf{P}}} = [\hat{\hat{P}}_x(RX)$, $\hat{\hat{P}}_y(RX)]$ represents the final estimated position, with $x$ and $y$ coordinates, of device/user $RX$, with $RX=\{1,2....N\}$ and $N$ the number of devices. $\hat{P}^{lat_2}_{x}(RX)$ and $\hat{P}^{lat_2}_{y}(RX)$ are the position estimated, $x$ and $y$ coordinates, with the collaborative lateration and $\hat{P}^{lat_1}_{x}(RX)$ and $\hat{P}^{lat_1}_{y}(RX)$ are the position estimated, $x$ and $y$ coordinates, with the stand-alone lateration, both for device/user $RX$.

\subsubsection{Full collaborative workflow}
Finally, algorithm \ref{alg:two} shows the pseudo-code for the collaborative positioning algorithm. The algorithm inputs are the shared information exchange by each collaborative device/user within a time window ($wt$) of \SI{60}{\second} and correspond to the \ac{rss} transmitted by each collaborative device/user ($RSS_{RX}(TX)$), position of each collaborative device/user estimated by the stand-alone lateration ($\hat{P}_{x}^{Lat_1}(TX) $,$\hat{P}_{y}^{Lat_1}(TX) $,$\hat{P}_{x}^{Lat_1}(RX) $,$\hat{P}_{y}^{Lat_1}(RX) $), and the \ac{rss} at \SI{1}{\meter} ($RSS^{1m}_{RX}(BX)$).
The algorithm workflow is summarized as follows:

\begin{itemize}
     
    \item \textbf{1\textsuperscript{nd} step:} Group the \ac{rss} readings by device and remove outlier values. We consider those values falling out of 25 and 75 percentiles (lines 1--2 in Algorithm~\ref{alg:two}) as outliers;
    
     \item \textbf{2\textsuperscript{rd} step:} Apply the average to the \ac{rss} values of each device/user, getting one averaged \ac{rss} value per device/user (line 3 in Algorithm~\ref{alg:two});

     \item \textbf{3\textsuperscript{th} step:} Estimate the relative distances of neighboring devices/users to the target device/user, using the \ac{mlp} model, which structure is depicted in Figure~\ref{fig:Ann_model}. The inputs of \ac{ann} model are:$RSS_{RX}(TX)$,  ($\hat{P}_{x}^{Lat_1}(TX) $,$\hat{P}_{y}^{Lat_1}(TX) $,$\hat{P}_{x}^{Lat_1}(RX) $,$\hat{P}_{y}^{Lat_1}(RX) $), $RSS^{1m}_{RX}(BX)$ and  (input values in Algorithm~\ref{alg:two});
    
    \item \textbf{4\textsuperscript{th} step:} Estimate the device/user's position ($\hat{\textbf{P}}^{Lat2}(RX)$) using the Levenberg-Marquardt Weighted Least Squares (L-MWLS) lateration method to fit the Euclidean Distance model. The input data to fit the model are the relative distances estimated by the \ac{mlp} neural network model (step 3), the weights and the estimated positions ($\hat{\textbf{P}}^{Lat1}$) of the neighboring devices/users. As in the stand-alone algorithm, the weight value for every \ac{ble} anchor is computed as the inverse of its distance square with respect to the device/user; 
    
    \item \textbf{5\textsuperscript{th} step:}
    Compute the final estimated device/user's position ($\hat{\hat{\textbf{P}}}(RX)$) based in the formula expressed in eq.\eqref{eqn:joinequation}, using the estimated standalone position $\hat{\textbf{P}}^{lat_1}(RX)$ and collaborative estimated position $\hat{\textbf{P}}^{lat_2}(RX)$, where $(RX)$ is the identifier of the device/user to estimate its position.
\end{itemize}
\begin{algorithm}[!htb]
 \caption{Collaborative module}
 \label{alg:two}
 \begin{algorithmic}[1]
 \small
 \renewcommand{\algorithmicrequire}{\textbf{Input:}}
 \renewcommand{\algorithmicensure}{\textbf{Output:}}
 \REQUIRE Collaborative devices information collected within a time window $tw$: $RSS_{RX}(TX)$, $\hat{P}_{x}^{Lat_1}(TX) $,$\hat{P}_{y}^{Lat_1}(TX) $,$\hat{P}_{x}^{Lat_1}(RX) $,$\hat{P}_{y}^{Lat_1}(RX) $,  and $RSS^{1m}_{RX}(BX)$
 \ENSURE Improved estimated device/user position ($\hat{\hat{\textbf{P}}}_{dev}(n)$)
 \STATE Group the $RSS_{RX}(TX)$ values by device
\STATE Remove $RSS_{RX}(TX)$ outliers values of each group
 \STATE Average $RSS_{RX}(TX)$ values of each group: $\overline{RSS_{dev}}(i)$
  \STATE Estimate the relative distance between the target device and the near collaborative devices using the trained \ac{ann} model
 \STATE Estimate the device/user's position ($\hat{\textbf{P}}^{Lat_2}(RX)$) using the Levenberg-Marquardt Weighted Least Squares (L-MWLS) Lateration method 
 \STATE Compute the final estimated device/user's position ($\hat{\hat{\textbf{P}}}(RX)$) using the  midpoint line algorithm of eq.\eqref{eqn:joinequation}
 \end{algorithmic}
 \end{algorithm}

\begin{table*}[!hbt]
\caption{Position of the anchors, BLE beacons and collaborative devices in the 7 configurations}
\label{tab_beaconlist}
\centering
\tabcolsep 4.75pt
\begin{tabular}{ccc|ccccccccccccccccccccc}
\toprule

 & \multicolumn{2}{c}{\textbf{BLE anchors}} && \multicolumn{2}{c}{\textbf{Config. 1}} && \multicolumn{2}{c}{\textbf{Config. 2}} && \multicolumn{2}{c}{\textbf{Config. 3}} && \multicolumn{2}{c}{\textbf{Config. 4}} && \multicolumn{2}{c}{\textbf{Config. 5}} && \multicolumn{2}{c}{\textbf{Config. 6}} && \multicolumn{2}{c}{\textbf{Config. 7}}\\ \cmidrule{2-3} \cmidrule{5-6} \cmidrule{8-9} \cmidrule{11-12} \cmidrule{14-15} \cmidrule{17-18} \cmidrule{20-21} \cmidrule{23-24}
 
 $n$& \textbf{x} & \textbf{y} && \textbf{x} & \textbf{y} && \textbf{x} & \textbf{y} &  & \textbf{x} & \textbf{y} && \textbf{x} & \textbf{y} && \textbf{x} & \textbf{y} && \textbf{x} & \textbf{y} && \textbf{x} & \textbf{y}\\ \midrule
1 & 0 & 0 && 5.05 & 3.7 && 1.33 & 6.1 && 6.93 & 1.3 && 7.75 & 6.1 && 2.05 & 9.7 && 2.05 & 9.7 && 2.05 & 9.7\\
2 & 0 &10.68 && 6.55 & 4.55 && 4.49 & 3.05 && 9.93 & 1.3 && 11.75 & 2.75 && 3.6 & 3.3 && 8.7 & 6.4 && 14.66 & 6.45\\
3 & 3.78 &6.51 &&  8.05 & 0.7 &&  7.66 & 0.1 && 12.93 & 1.3 && 12.75 & 0.1 && 16.45 & 2.5 && 16.45 & 2.5 && 16.45  & 2.5\\
4 & 6.68  &10.64 && 5.05 & 0.7 && 1.33 & 0.1 && 9.03 & 0.1 && 7.75 & 0.1 && 2.05 & 2.5 && 2.05 & 2.5 && 2.05 & 2.5\\
5 & 9.2 &3.7 && 8.05 & 3.7  && 7.66 & 6.1  && 9.03 & 3.7 && 12.75 & 6.1 && 16.45 & 9.7 && 16.45 & 9.7 && 16.4 & 9.7\\
6 & 14.2 &6.05 && - & -  && - & -  && - & - && - & - && - & - && - & - && - & -\\
7 & 16.65 & 10.65 && - & -  && - & -  && - & - && - & - && - & - && - & - && - & -\\

\bottomrule
\end{tabular}
\end{table*}

\section{Experiments and Results}
\label{sec:experiments}
\subsection{Objectives and Experimental setup}
The principal goal of our experiments is to assess the feasibility and benefits of our collaborative approach with respect to a traditional \ac{ble}-\ac{rss} lateration baseline to mitigate inadequate deployment of \ac{ble} anchors and increased \ac{nlos} conditions. Specifically, we evaluate and compare the positioning accuracy of both in a realistic environment.

\begin{figure}[!h]
    \centering
    \includegraphics[width=1\columnwidth]{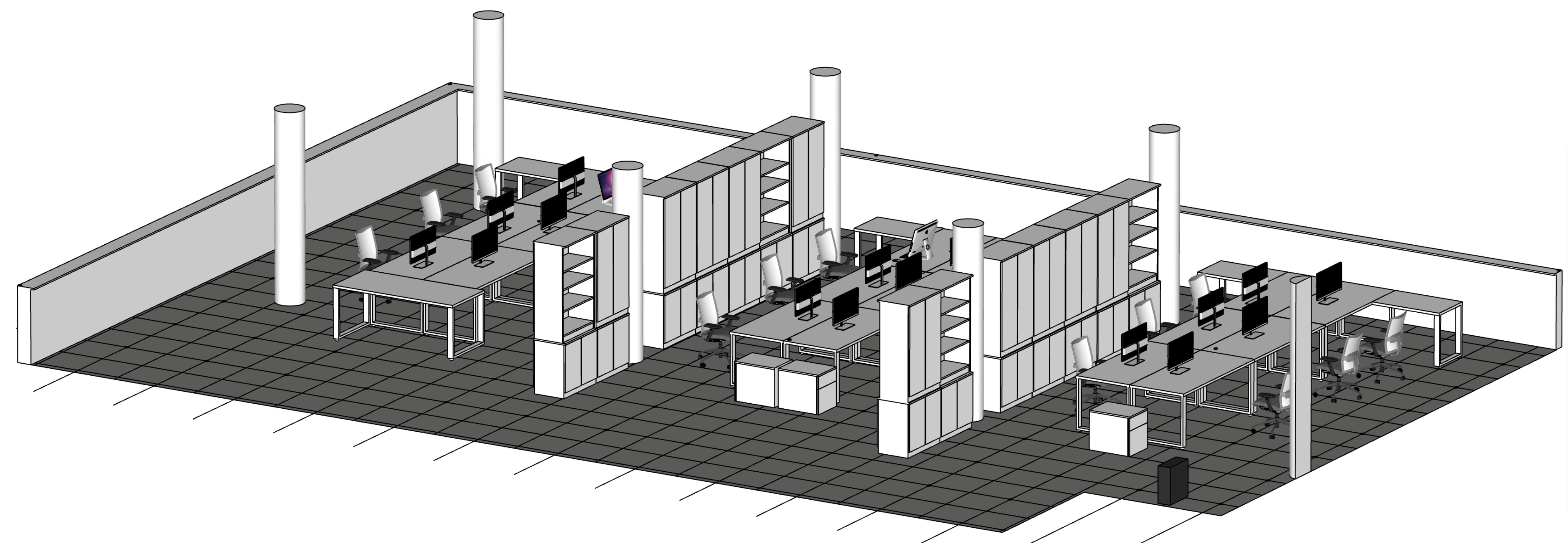}
  \caption{3D Office Scenario representation}
  \label{fig:3DScenario}
\end{figure}

We performed the experiments in a real office scenario. The office covers an approximate area of \SI[product-units=power]{10.8 x 16.7}{\meter} and contains seven deployed \ac{ble} beacons (anchors), with a transmission power and period of \SI{-4}{\dBm} and  \SI{250}{\milli\second} respectively. Figure \ref{fig:3DScenario} shows the complete scenario through a 3D representation, which allows us to illustrate the complexity of the environment. The  \ac{nlos} conditions are mainly created by the furniture, which includes chairs, desks, desktop computers, bookshelves, and the concrete metal-reinforced pillars.  

In this office scenario, a comprehensive data collection was conducted. First, the five collaborative devices were registered during the registration phase (see Section~\ref{sec:registering}) and their $RSS^{1m}_{RX}$ was measured using a fixed \ac{ble} beacon from the environment. Table~\ref{rssat1mtable} summarizes the five devices that were used in the experiment, and their corresponding $RSS^{1m}_{RX}$ value, which depend on the smartphone model and range from \SIrange{-78.79}{-62.39}{\dbm}.

\begin{table}[htbp]
\caption{$RSS^{1m}_{RX}(BX)$ values by device}
\label{rssat1mtable}
\centering
\begin{tabular}{ccc}
\toprule
$RX$ & Device & $RSS^{1m}_{RX}(BX)$ values (\si{\dBm})\\ 
\midrule
1 & Galaxy S8 & -68.88  \\
2 & Lenovo Yoga Book & -74.75  \\
3 & Galaxy A7 Duos & -62.39  \\
4 & Galaxy S6 & -62.99  \\
5 & Galaxy A5 & -78.79 \\
\bottomrule
\end{tabular}
\end{table}

\begin{figure}[!h]
  \centering
  \includegraphics[width=1\linewidth]{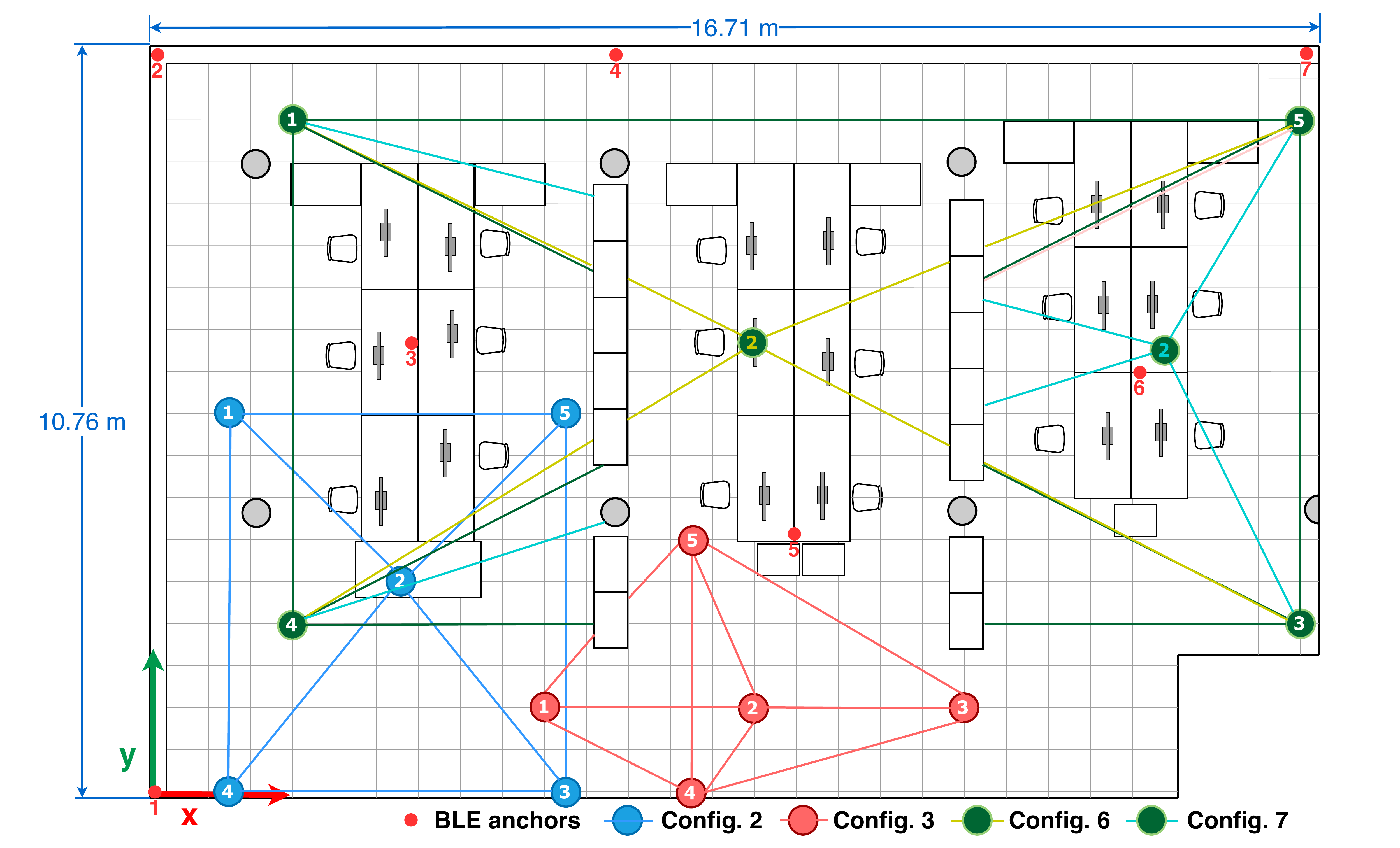}
  \caption{Distribution of the configuration 2, 3, 6, and 7 used in the training \& validation and distribution of the \ac{ble} anchors}
  \label{fig:sketch_val_train}
\end{figure}

\begin{figure}[!h]
  \centering
  \includegraphics[width=1\linewidth]{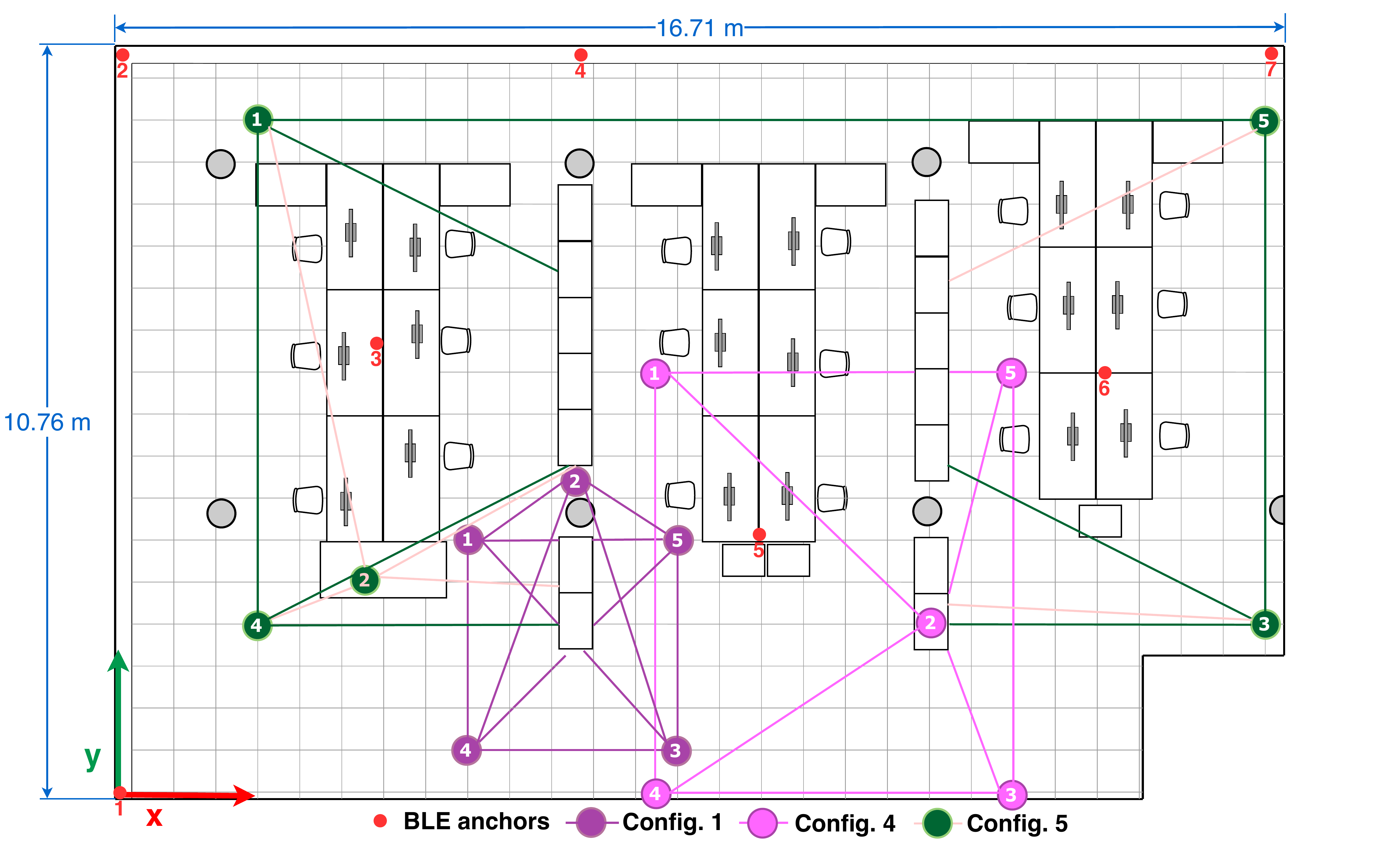}
  \caption{Distribution of the configuration 1, 4, and 5 used in the testing and distribution of the \ac{ble} anchors}
  \label{fig:sketch_test}
\end{figure}

Then, seven different collaborative distributions (called configurations in this paper) made up of these five devices were considered. In each configuration, the five devices simultaneously broadcast and save the information of surrounding devices, including the \ac{ble} anchors, for 2 hours. Devices broadcast every 100 ms in accordance with the low latency mode supported by Android devices~\cite{kindt2021reliable}. However, as mobile devices are not ad-hoc positioning devices, the broadcast period is affected mainly by the priority of executing tasks and power saving modes implemented by operating systems. The data collection for each configuration was performed independently, at a different time. As already mentioned, for the lateration approach based on the \acf{ldpl}, the path-loss factor was determined to be $\eta = 2.1$ for all the five devices. Figure \ref{fig:sketch_val_train} and Figure \ref{fig:sketch_test} show the distribution of the devices in each configuration and the \ac{ble} anchors in the scenario, and  Table~\ref{tab_beaconlist} summarizes the \ac{gt} coordinates of them.

\begin{figure*}
\centering
\tabcolsep 0pt
\begin{tabular}{cccc}
\includegraphics[clip, trim=2.6cm 6.5cm 3cm 6.5cm,width=0.25\textwidth]{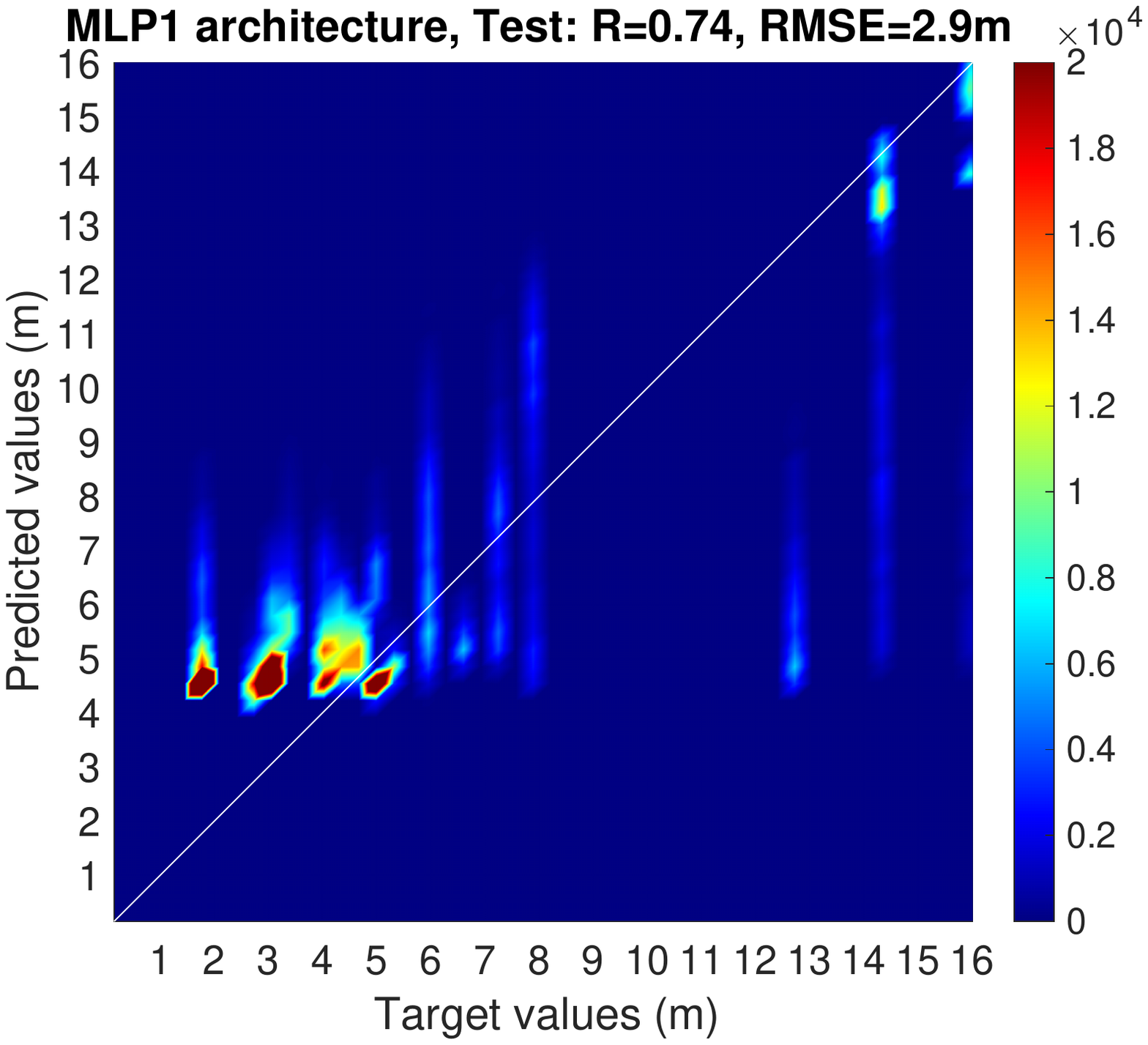}&
\includegraphics[clip, trim=2.6cm 6.5cm 3cm 6.5cm,width=0.25\textwidth]{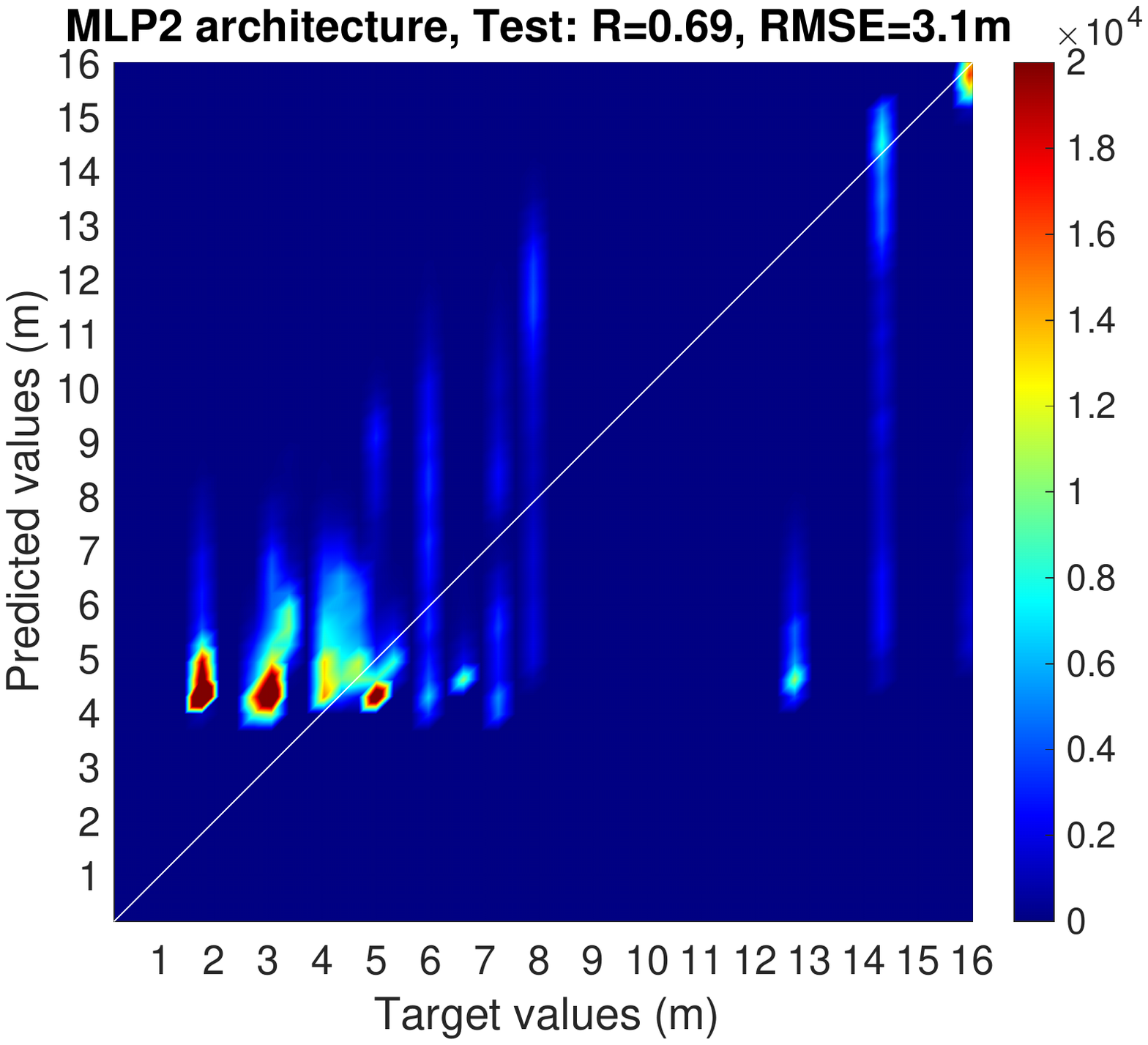}&
\includegraphics[clip, trim=2.6cm 6.5cm 3cm 6.5cm,width=0.25\textwidth]{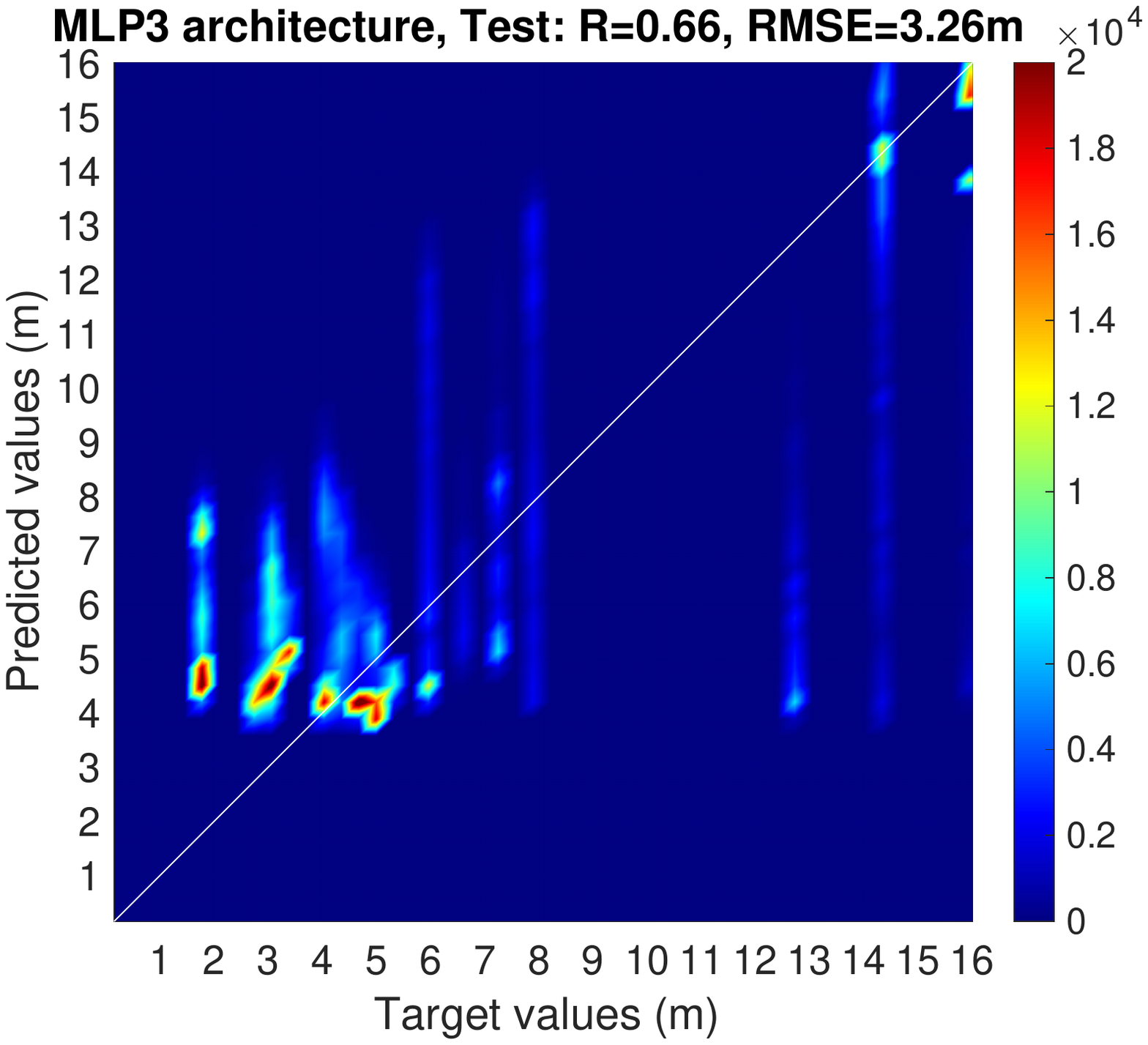}&
\includegraphics[clip, trim=2.6cm 6.5cm 3cm 6.5cm,width=0.25\textwidth]{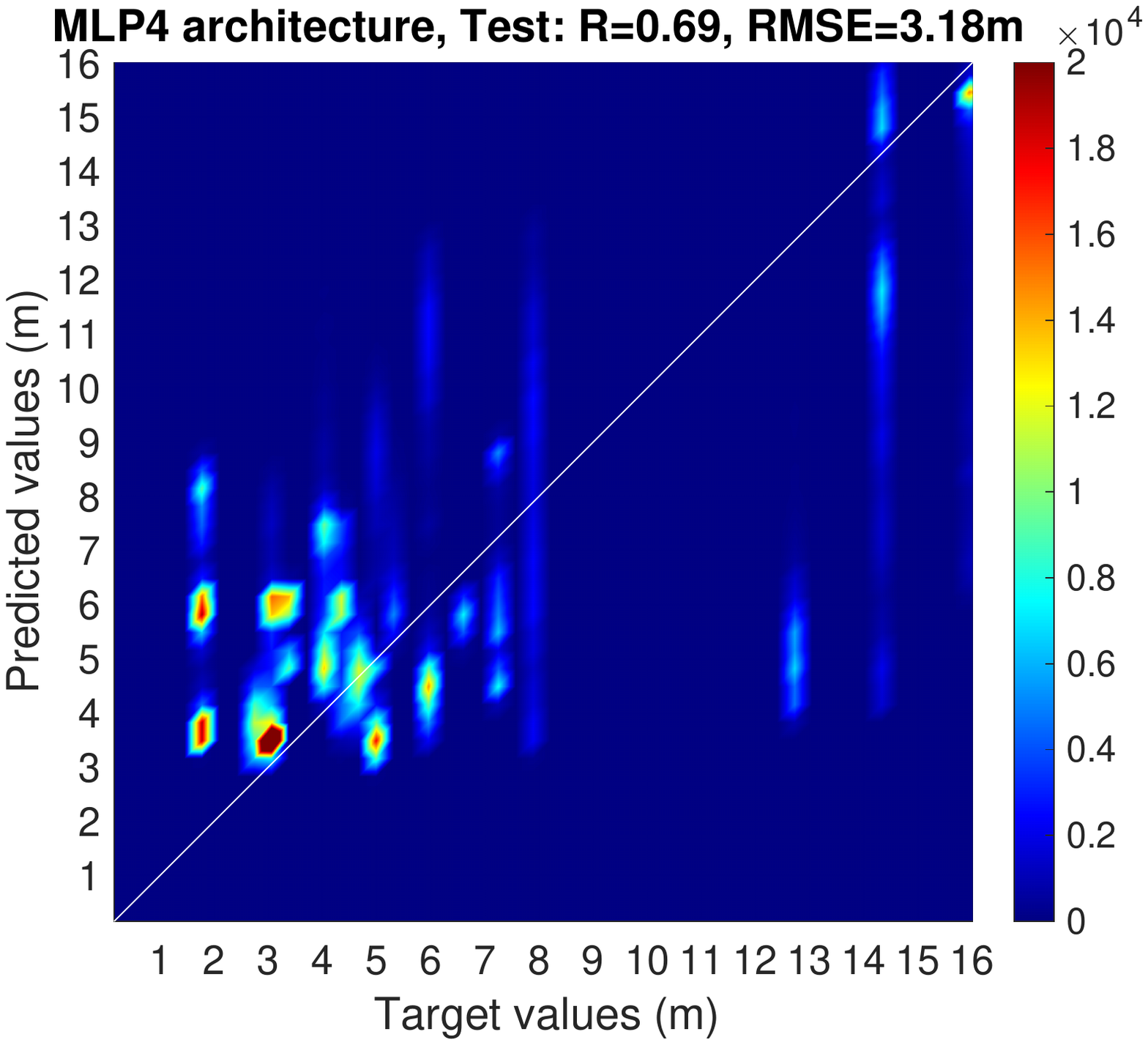}\\
(a) \ac{mlp}1&(b) \ac{mlp}2&(c) \ac{mlp}3&(d) \ac{mlp}4\\
\end{tabular}

\caption{Target vs Predicted distances from the test dataset estimated with \ac{mlp}1--\ac{mlp}4 architectures}
\label{fig:ann1234}
\end{figure*}

The data collected in the seven configurations were split into two datasets, where the devices cover different areas, and the distances among devices, and between devices and anchors, varies. Each dataset is used for different purposes: configurations 2, 3, 6, and 7 (see Fig.~\ref{fig:sketch_val_train}) were used to train the model and configurations 1, 4, and 5 (see Fig.~\ref{fig:sketch_test}) were used for evaluation. In order to tune the \ac{mlp} neural network, the first dataset was randomly split into training ($70\%$) and validation ($30\%$) with $968798$ and $242199$ samples respectively. The testing dataset contains $936103$ samples and it was used to test the \ac{mlp} \ac{ann}  for relative distances, and implement the \ac{ble}-\ac{rss} lateration baseline and collaborative approach. 

The selected scenario is challenging and covers an area with low \ac{gdop} and strong heterogeneous \ac{nlos} conditions (diversity in obstacles): only 7 \ac{ble} beacons provide positioning, and there is device diversity in the collaborative phase.
The proposed collaborative model targets dynamically extending the anchor coverage by using the collaborative user's devices, replacing the traditional \ac{ldpl} model with a \ac{mlp} \ac{ann} that processes the \ac{rss} considering the receiver and emitter location and the calibration of \ac{rss} (at \SI{1}{\meter}) to minimise the device diversity effect. 

\subsection{\acf{mlp} neural network tuning}
We tested four \acf{mlp} neural network architectures in order to determine the best architecture and set of hyperparameters of our model. As hyperparameters, we considered different numbers of hidden layers (1 and 2) and  neurons, as well as two types of activation functions, the Hyperbolic tangent sigmoid (tansig) and Log-sigmoid (logsig). Table~\ref{tab_hyperparameters} summarizes the hyperparameters configured in the proposed architectures.

\begin{table}[htbp]
\caption{Tested \ac{mlp} architectures and hyper parameters}
\label{tab_hyperparameters}
\centering
\begin{tabular}{lcccc}
\toprule
Parameters & MLP1 & MLP2 & MLP3 & MLP4 \\ 
\midrule
No. Input layers & 6 & 6 & 6 & 6 \\
No. Hidden layers (HLs)& 1 & 1 & 2 & 2 \\
No. Output Layer & 1 & 1 & 1 & 1 \\
No. Neurons HL1  & 3 & 3 & 6 & 12 \\
No. Neurons HL2  & - & - & 3 & 6 \\
Training function & \multicolumn{4}{c}{trainscg}\\ 
Activation function & tansig & logsic & tansig & tansig\\
Performance Function & \multicolumn{4}{c}{Mean Square error} \\
\bottomrule
\end{tabular}
\end{table}

Before presenting the main positioning results of the proposed collaborative approach and the lateration baseline, we present the results of evaluating the four proposed \ac{mlp} architectures with the training dataset in order to determine which of them provides a more accurate estimation of relative distances. Figure~\ref{fig:ann1234} shows the density scatter plots of the real distance against estimated distance by each \ac{mlp} neural network architecture. Also, the \ac{rmse} and correlation coefficient (R) values are indicated in each of them. 

From those values and density scatter plots, we can notice that the \acf{mlp} architectures with one hidden layer (\ac{mlp}1 and \ac{mlp}2) present a \ac{rmse} lower than the architectures with two hidden layers (\ac{mlp}3 and \ac{mlp}4), as well as a greater correlation coefficient ($0.74$ and $0.69$ for \ac{mlp}1 and \ac{mlp}2 respectively). Therefore, architectures with a single hidden layer are able to estimate the distance more accurately. In specific, \ac{mlp}1, which use an hyperbolic tangent sigmoid activation function (tansig),  presents a higher density of predicted values closer to the real values in comparison with the \ac{mlp}2, which use a Log-sigmoid activation function, as can be observed in the density scatter plots of Figure~\ref{fig:ann1234} (a) and (b). Additionally, we notice that the density increases for the range between \SIrange{3}{6}{\meter} and around \SI{15}{\meter}. The results of high density values around the short distance demonstrate the potential of the \ac{mlp} architecture to improve the accuracy of position estimations in collaborative \ac{ips} in scenarios with a high density of collaborating mobile devices (i.e., when the distance between neighboring devices is short).

As a result of assessing different \ac{mlp} architectures, the \ac{mlp}1  with one hidden layer (3 neurons), a scaled conjugate gradient backpropagation training function, and a hyperbolic tangent sigmoid activation function was selected to estimate relative distances with \ac{ble}--\ac{rss} data.

\subsection{Results of the collaborative model}

Table~\ref{tab_targetestimation} presents the main results of the lateration baseline and our collaborative approach using various evaluation metrics (i.e., \ac{rmse}, mean, median, 70th and 90th percentile), and also indicates the relative difference between them. 

\begin{table}[htbp]
\caption{Main results metrics provided by the Lateration baseline and our proposed collaborative approach}
\label{tab_targetestimation}
\centering
\aboverulesep = 0.15125mm
\begin{tabular}{cScSSc}
\toprule
&
\multicolumn{1}{c}{Lateration baseline} &&
\multicolumn{2}{c}{Collaborative approach}
 \\\cmidrule{2-2}\cmidrule{4-5}
\multicolumn{1}{c}{Eval. metric} & 
\multicolumn{1}{c}{Error (m)} &&
\multicolumn{1}{c}{Error (m)} & 
\multicolumn{1}{c}{Diff.} \\
\midrule
RMSE & 5.76 && 5.22 & $\downarrow$ 9.37\% \\
Mean & 5.36 && 4.55 & $\downarrow$ 15.11\% \\
Median & 5.54 && 4.84 & $\downarrow$ 12.63\% \\
$75^{th}$ percentile & 7.08 && 6.68 & $\downarrow$ 5.64\% \\
$90^{th}$ percentile & 8.08 && 8.08 & 0\% \\
\bottomrule
\end{tabular}
\label{tab:results}
\end{table}

As reported in Table~\ref{tab_targetestimation}, our collaborative approach outperforms the lateration baseline in four of the five evaluation metrics, with a maximum of $15.11\%$ of difference for the ``mean'' metric.

\begin{figure}[!h]
    \centering
  \includegraphics[width=0.85\linewidth]{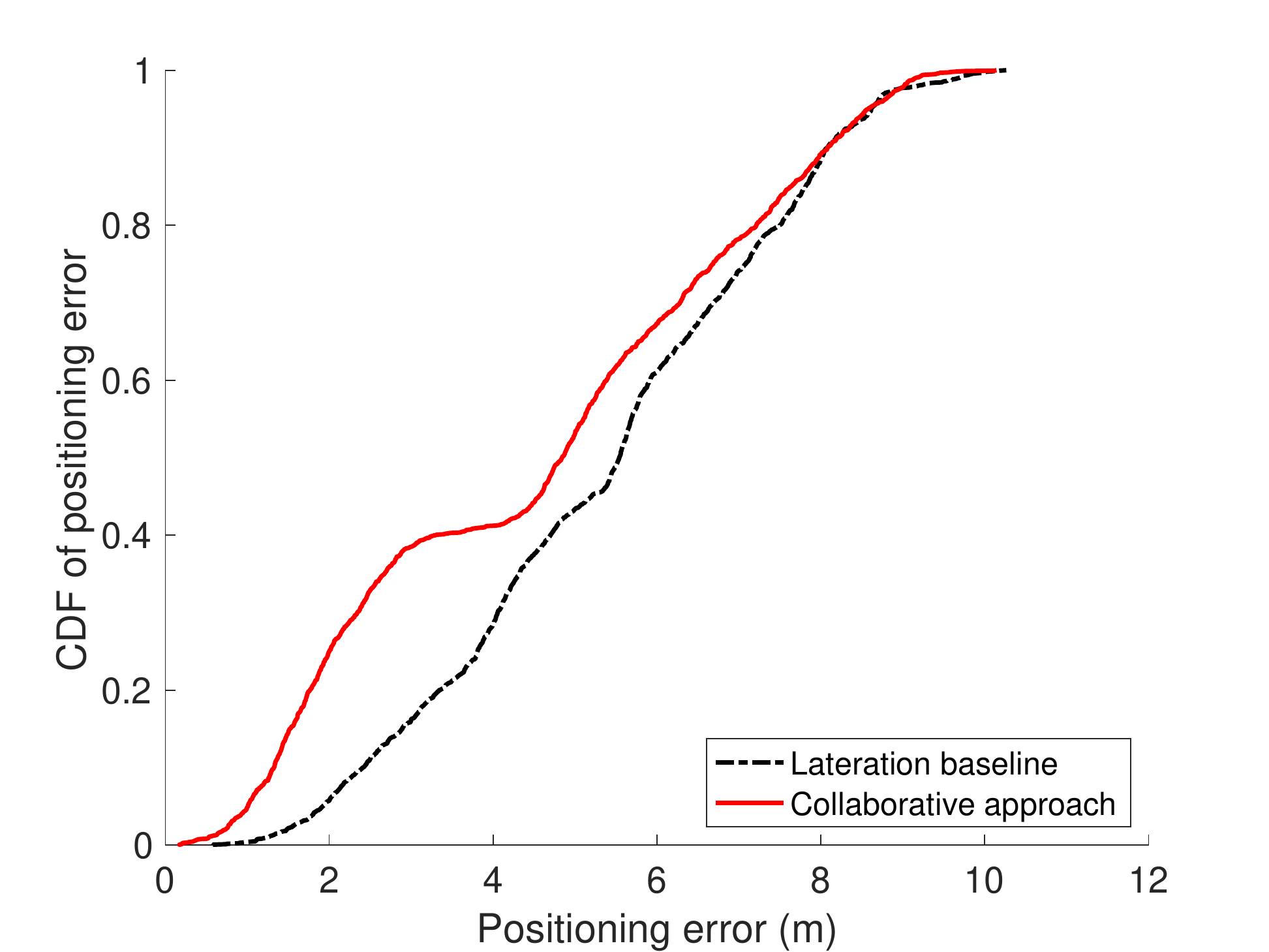}
  \caption{Empirical CDF provided by the Lateration baseline and  our Collaborative approach}
  \label{fig:cdf_plot}
\end{figure}

 Figure~\ref{fig:cdf_plot} introduces the \ac{ecdf} of the lateration baseline (black dotted line) and our proposed approach (red line). For the first 40\% of cases, our collaborative model significantly outperforms the baseline, for the next $50\%$ of cases, the improvement is moderate, being quite close for the last $10\%$ of cases. 
 
It should be noted that the positioning accuracy of the lateration baseline was significantly degraded, compared to BLE-RSS approaches reported in the literature (\SIrange{2}{3}{\meter} error \cite{pascacio2021lateration,subhan2019experimental}), due to the adverse test scenario conditions, namely poor quantity and inappropriate distribution of anchors, unstable signal strength and hardware heterogeneity. However, the magnitude of the errors is in line with scenarios that consider the low-density anchor deployment issue. For example, \textcite{cengiz2020comprehensive} reported a mean error accuracy of \SI{7.5}{\meter} in their lateration algorithm, which was tested on a \SI[product-units=power]{24 x 24}{\meter} area with 8 anchors.

\begin{figure}[!h]
  \centering
  \includegraphics[width=0.9\linewidth]{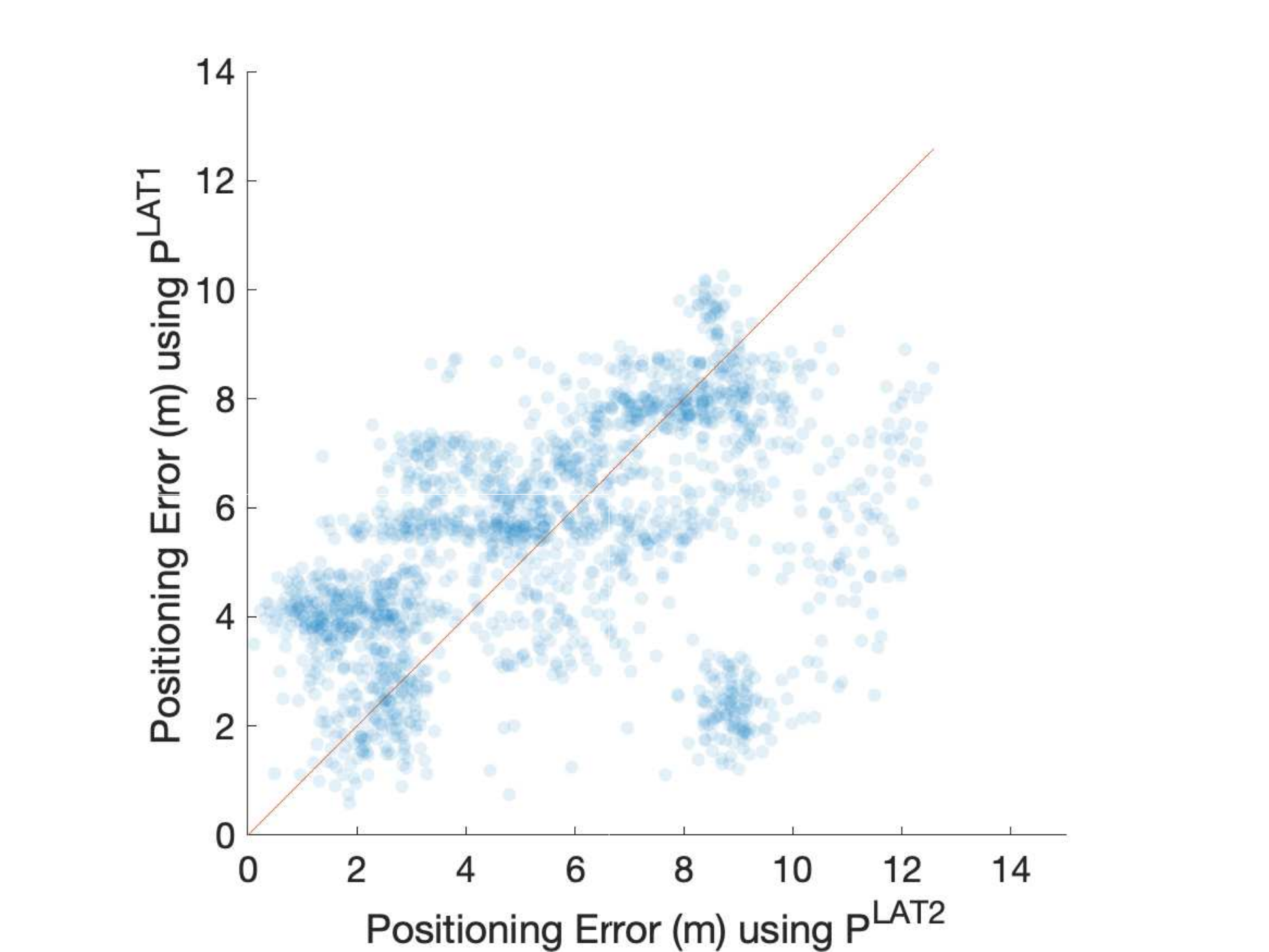}
  \caption{Comparison of individual errors provide by the Lateration baseline and  our Collaborative lateration approach}
  \label{fig:scaterplot}
\end{figure}

Figure~\ref{fig:scaterplot} provides a more detail representation of the individual positioning errors of the laterations used in the stand-alone phase and the collaborative phase.

From the scatter plot, we can deduce that the lateration used in the calibration phase is providing better positioning results than the one based on the few beacons deployed in the environment. However, for a few cases the behavior is just the contrary as the accuracy is a few meters worse than the baseline. Nevertheless, the strongest feature of the proposed collaborative model is the combination of the lateration results from both phases, providing an ensembled output. According to the ensemble theory, the combination of estimators improves the accuracy of any of the individuals involved. 

\section{Conclusion and Future Work}
\label{sec:conclusions}
In this paper, we presented a collaborative indoor positioning system using \acf{mlp} \acfp{ann} to enhance the accuracy of the models based on \ac{ble} and \ac{rss} lateration, in typical adverse conditions, namely poor quantity and inappropriate distribution of anchors, unstable signal strength and hardware heterogeneity. The proposed approach is divided in three phases. The first phase is devoted to registering devices used in the collaborative approach and establishing an \ac{rss} baseline measurement at \SI{1}{\meter}, the second phase consists of the stand-alone (non-collaborative) lateration algorithm for position estimation of each device/user, and the last phase is devoted to collaboratively estimate the position of the target device/user and combine it with the non-collaborative position estimate. 

We evaluated our collaborative system in an indoor scenario (Office) and compared it with a lateration baseline. Our indoor scenario presents the aforementioned typical adverse conditions: an inadequate deployment of \ac{ble} anchors, rich \ac{nlos} conditions and heterogeneity among mobile devices used, causing a considerably reduction of positioning accuracy in \ac{rss}-based approaches. Experimental results show the feasibility and benefits of our proposed collaborative approach to outperform the positioning accuracy of the traditional lateration baseline under these adverse conditions. Particularly, our approach decreases, with respect to the lateration baseline, the mean, median, \ac{rmse} and  $75^{th}$ percentile positioning error metrics with $15.11\%$, $12.63\%$, $9.37\%$ and $5.64\%$ respectively. 

Furthermore, the results highlighted the usefulness of an \ac{mlp} neural network model to model the signal propagation at short distance considering the inherent characteristic of each receiving device and in \ac{nlos} conditions, which is essential for estimating the distance between (heterogeneous) collaborative devices in real-world conditions.

As future work, we plan to use robust baseline approaches, which are able to provide a better positioning accuracy. We will consider more complex algorithms to combine independent collaborative estimates and reduce the overall positioning error. Additionally, we will enhance the neural network tuning, testing it in diverse and complex scenarios.

\balance
\renewcommand*{\UrlFont}{\rmfamily}
\printbibliography

\end{document}